\documentclass[dvipdfmx]{pasj01}
\draft
\Received{$\langle$reception date$\rangle$}
\Accepted{$\langle$acception date$\rangle$}
\Published{$\langle$publication date$\rangle$}
\usepackage{mathpazo}
\usepackage[T1]{fontenc}

\begin{document}

\title{ {\LARGE LETTER} \\
Discovery of a distant molecular cloud in the extreme outer Galaxy with the Nobeyama 45-m telescope}
\author{Mitsuhiro \textsc{Matsuo},\altaffilmark{1, 2, *}
Hiroyuki \textsc{Nakanishi},\altaffilmark{1, 3}
Tetsuhiro \textsc{Minamidani},\altaffilmark{2, 4}
Kazufumi \textsc{Torii},\altaffilmark{2}
Masao \textsc{Saito},\altaffilmark{2, 4}
Nario \textsc{Kuno},\altaffilmark{5}
Tsuyoshi \textsc{Sawada},\altaffilmark{6, 7}
Tomoka \textsc{Tosaki},\altaffilmark{8}
Naoto \textsc{Kobayashi},\altaffilmark{9, 10}
Chikako \textsc{Yasui},\altaffilmark{11}
Hiroyuki \textsc{Mito},\altaffilmark{10}
Takashi \textsc{Hasegawa},\altaffilmark{12}
and 
Akihiko \textsc{Hirota}\altaffilmark{6, 7}
}
\altaffiltext{1}{Graduate School of Science and Engineering, Kagoshima University, 1-21-35 Korimoto, Kagoshima, Kagoshima 890-0065, Japan}
\altaffiltext{2}{Nobeyama Radio Observatory, National Astronomical Observatory of Japan, National Institutes of Natural Sciences, 462-2 Nobeyama Minamimaki, Minamisaku, Nagano 384-1305, Japan}
\altaffiltext{3}{Institute of Space and Astronautical Science, Japan Aerospace Exploration Agency, 3-1-1 Yoshinodai, Chuo-ku, Sagamihara, Kanagawa 252-5210, Japan}
\altaffiltext{4}{Department of Astronomical Science, School of Physical Science, SOKENDAI (The Graduate University of Advanced Studies), 2-21-1 Osawa, Mitaka, Tokyo 181-8588, Japan}
\altaffiltext{5}{Graduate School of Pure and Applied Sciences, University of Tsukuba, 1-1-1 Tennodai, Tsukuba, Ibaraki 305-8577, Japan}
\altaffiltext{6}{Joint ALMA Office, Alonso de C\'{o}rdova 3107, Vitacura, Santiago 763-0355, Chile}
\altaffiltext{7}{NAOJ Chile Observatory, Joaqu\'{i}n Montero 3000 Oficina 702, Vitacura, Santiago 763-0409, Chile}
\altaffiltext{8}{Joetsu University of Education, Yamayashiki-machi, Joetsu, Niigata 943-8512, Japan}
\altaffiltext{9}{Institute of Astronomy, School of Science, University of Tokyo, 2-21-1 Osawa, Mitaka, Tokyo 181-0015, Japan}
\altaffiltext{10}{Kiso Observatory, Institute of Astronomy, School of Science, University of Tokyo, 10762-30 Mitake, Kiso-machi, Kiso-gun, Nagano 397-0101, Japan}
\altaffiltext{11}{National Astronomical Observatory of Japan, 2-21-1 Osawa, Mitaka, Tokyo 181-8588, Japan}
\altaffiltext{12}{Gunma Astronomical Observatory, 6860-86 Nakayama, Takayama, Agatsuma, Gunma 377-0702, Japan}

\email{k2330611@kadai.jp}

\KeyWords{ISM: clouds --- ISM: molecules ---radio lines: ISM}

\maketitle

\begin{abstract}
We report the discovery of the molecular cloud whose kinematic distance is the largest in the Galaxy at the present moment, named G213.042$+$0.003, at $l =$ \timeform{213.042D} and $b =$ \timeform{0.003D} in the \atom{C}{}{12}\atom{O}{}{}($J =$ 1--0) line using the Nobeyama 45-m telescope and a multi-beam receiver BEARS.
This molecular cloud is located at the heliocentric distance of 21$_{-7}^{+12}$~kpc and Galactocentric distance of 29$_{-7}^{+12}$~kpc, which are estimated as the kinematic distances with the Galactic parameters obtained by Reid et al. (2014, ApJ, 783, 130).
Its major and minor diameters and line width were measured to be 4.0$_{-1.3}^{+2.3}$~pc, 3.0$_{-1.0}^{+1.7}$~pc, and 1.5~km~s$^{-1}$, respectively.
The cloud mass was estimated to be 2.5$_{-1.4}^{+3.7}$ $\times$ 10$^2$~$\Mo$ using the \atom{C}{}{}\atom{O}{}{}-to-\atom{H}{}{}$_2$ conversion factor of 5.6 $\times$ 10$^{20}$~cm$^{-2}$~(K~km~s$^{-1}$)$^{-1}$ obtained in far outer Galaxy.
\end{abstract}

\section{Introduction}
The latest observational studies show that many spiral galaxies have extended ultraviolet (XUV) disks with radii a couple of times larger than optical disks.
Such outermost disks were found in many galaxies at the redshift range from $z =$ 0.001 to $z =$ 0.05 by studying {\it Galaxy Evolution Explorer} ({\it GALEX}) deep imaging data and Sloan Digital Sky Survey DR7 footprints \citep{2011ApJ...733...74L}. 

Observational study of molecular clouds in the outermost disks such as XUV disks is essential to reveal their origin, because the UV emission is mainly emitted from young massive stars and those stars are formed in interstellar molecular clouds.
Since the molecular-hydrogen surface density ($\Sigma_\mathrm{H_2}$) of the outermost disks decreases to less than 1~$\Mo$~pc$^{-2}$, \citet{2016MNRAS.455.1807W} analyzed the deepest \atom{C}{}{}\atom{O}{}{} data for outermost disk of NGC 4625 down to 0.11~$\Mo$~pc$^{-2}$ with a 1~kpc beam.
However, the \atom{C}{}{}\atom{O}{}{} emission was not detected at the outermost region where star formation activities are shown and the \atom{H}{}{}~{\sc i} surface density ($\Sigma_\mathrm{HI}$) of $\sim$ 2~$\Mo$~pc$^{-2}$.
One of the reasons why \atom{C}{}{}\atom{O}{}{} emission was not detected is that observations for external galaxies cannot reach enough spatial resolution and sensitivity.

The Milky Way Galaxy is the best target for finding molecular clouds in the outermost disk since it is the closest galaxy and molecular clouds are brighter and can be observed with better spatial resolution than any external galaxies.
Recent studies show that young stars are distributed beyond the Galactocentric distance of $R =$ 20~kpc (\cite{2008ASSP....5..315N}, \cite{2010ApJ...718..683C}).
Since these early-type stars are found in the outermost disk where $\Sigma_\mathrm{HI}$ decreases less than 1~$\Mo$~pc$^{-2}$, they are likely to form in the same environment as the outermost disk in external galaxies.

There have been many efforts to search for molecular clouds in the outermost disk of the Galaxy (e.g., \cite{1994ApJ...422...92D}, \cite{2015ApJ...798L..27S}).
\citet{1989A&AS...80..149W} found that molecular clouds are distributed up to $R =$ 20~kpc by observation towards {\it IRAS} sources.
However, an unbiased \atom{C}{}{}\atom{O}{}{} survey towards the outer Galaxy is essential, because there are many molecular clouds without {\it IRAS} sources. 
In the second Galactic quadrant, the molecular cloud at $R = 28$~kpc was detected \citep{1994ApJ...422...92D}.
Furthermore, the latest outer Galaxy survey with the 13.7~m telescope of the Purple Mountain Observatory (PMO) discovered a lot of molecular clouds associated with a new arm beyond the Outer (Cygnus) arm \citep{2015ApJ...798L..27S}.
However, no survey towards the third Galactic quadrant, where candidates of distant young stars found by \citet{2008ASSP....5..315N}, has yet been carried out.

In order to examine whether there exist distant molecular clouds in the third quadrant, we carried out wide field \atom{C}{}{12}\atom{O}{}{} mapping observation with the Nobeyama 45-m telescope.
The multi-beam receiver BEARS (25-BEam Array Receiver System; \cite{2000SPIE.4015..237S}) enables us to conduct a wide-field unbiased molecular survey with high spatial resolution by observing in On-The-Fly (OTF) mode \citep{2008PASJ...60..445S}. 
This letter reports a discovery of the molecular cloud (G213.042+0.003), whose kinematic distance is the largest ($R =$ 29$_{-7}^{12}$~kpc) to date.

\section{Observations}
We carried out mapping observations towards regions of $l =$ \timeform{165.750D} -- \timeform{166.250D}, \timeform{194.500D} -- \timeform{195.000D}, \timeform{197.500D} -- \timeform{198.500D}, \timeform{203.500D} -- \timeform{204.500D}, \timeform{208.167D} -- \timeform{209.167D}, and \timeform{212.500D} -- \timeform{214.000D} and $|b| <$ \timeform{0.25D} in the \atom{C}{}{12}\atom{O}{}{}($J =$ 1--0) (115.271204~GHz) line with the Nobeyama 45-m telescope and BEARS in the OTF mode between January and March of 2007.
In this letter, we focus on the region of $l =$ \timeform{212.5D} -- \timeform{214.0D} and $|b| <$ \timeform{0.25D}, where G213.042$+$0.003 was found.
The beam size of the telescope was \timeform{15"} and the main beam efficiency $\eta_{\rm MB}$ was 34 \% at 115 GHz.
Twenty-five BEARS beams were arrayed in a square of five rows and five lines with separation of \timeform{41".1}.
The beam array was oriented at an angle of seven degrees relative to the OTF scan direction so that all separations of the scan tracks were equal to five arcseconds.
The sampling interval along the scan path was three arcseconds.
We took two orthogonal  scan patterns in the galactic longitude and latitude directions in order to eliminate the scan pattern.
The received signal was converted to the intermediate frequency (IF) signal, which was sampled by a digital XF-type autocorrelator system (AC45; \cite{2000SPIE.4015...86S}) with band width of 512~MHz and 1024 channels.
The channel separations of frequency and velocity were 500~kHz and 1.3~km~s$^{-1}$ at 115~GHz, and velocity resolution was 1.6~km~s$^{-1}$ using rectangle window function.
The system noise temperature was typically 400~K in double sideband during the observations.
We converted output signal into antenna temperature $T_\mathrm{A}^\ast$ (K) to compensate for atmospheric attenuation using the chopper wheel method \citep{1976ApJS...30..247U}.
We checked whether the pointing accuracy was within five arcseconds by observing point-like SiO maser source, GX Monocerotis, every hour and a half.
We calibrated gain variation among 25 beams of BEARS by using the calibration table provided by Nobeyama Radio Observatory, which was measured by observing W51 with BEARS and the AC45 correlator and with the S100 receiver and acousto-optical spectrometers (AOSs) and by comparing them.

\section{Data analysis}
\subsection{Data reduction}
We made three-dimensional FITS cube using the software, Nobeyama OTF Software Tools for Analysis and Reduction (NOSTAR).
The procedure was as follows; (1) calibration of gain variation among 25 beams of BEARS, (2) subtraction of baseline by fitting with polynomial function, and (3) making the map of \timeform{15''} grid using the Bessel $\times$ Gauss function as the convolution function.
As a result the effective spatial resolution (FWHM) was \timeform{31.7''} \citep{2008PASJ...60..445S}.
The edges of observed areas were not included in the final cube because of their large noise level.
Since the root mean square (rms) noise level ($T_\mathrm{rms}$) was 0.22~K in $T_\mathrm{MB}$ scale, as small molecular cloud as 29~$\Mo$ at the heliocentric distance of $D =$ 20~kpc can be detected with three sigma level. 
For comparison, $T_\mathrm{rms}$ of outer Galaxy survey with FCRAO \citep{1998ApJS..115..241H} and PMO \citep{2015ApJ...798L..27S} were 0.73 and 0.18~K with a velocity width of 1.3~km~s$^{-1}$, respectively, corresponding $M_\mathrm{CO}$ of a molecular cloud at the heliocentric distance of 20~kpc were 2.0 $\times$ 10$^2$ and 64~$\Mo$, respectively.
Therefore, our \atom{C}{}{12}\atom{O}{}{} survey data enable us to detect more distant molecular clouds than ever.

\subsection{Cloud identification}
We used the DENDROGRAM algorithm developed by \citet{2008ApJ...679.1338R} to identify molecular clouds.
DENDROGRAM extracts a group of pixels satisfying the minimum contour level $T_\mathrm{min}$ and minimum pixel number $n_\mathrm{pix}$, which is called TRUNK.
Similarly, it extracts a group of pixels with larger value than $T_\mathrm{min}$ by $T_\mathrm{delta}$, which is called BRANCH.
In the same way, it extracts local maximum called LEAF from BRANCH.
We adopted parameters as $T_\mathrm{min} =$ 3$T_\mathrm{rms}$, $T_\mathrm{delta} =$ 2$T_\mathrm{rms}$, and $n_\mathrm{pix} =$ 4 pixels and we defined TRUNKs as molecular clouds in this letter.
Since this procedure sometimes misidentified noise as very small clouds, we excluded molecular cloud candidates with geometric mean of major and minor diameters smaller than the effective spatial resolution of \timeform{31.7''} and with line widths smaller than the velocity width of 1.3~km~s$^{-1}$.
Furthermore, we focus on molecular cloud candidates whose local standard of rest (LSR) velocity, $V_\mathrm{LSR}$, was within the range from 0 to 130~km~s$^{-1}$ so that the kinematic distance of clouds can be calculated.

\begin{figure}
 \begin{center}
  \includegraphics[width=100mm]{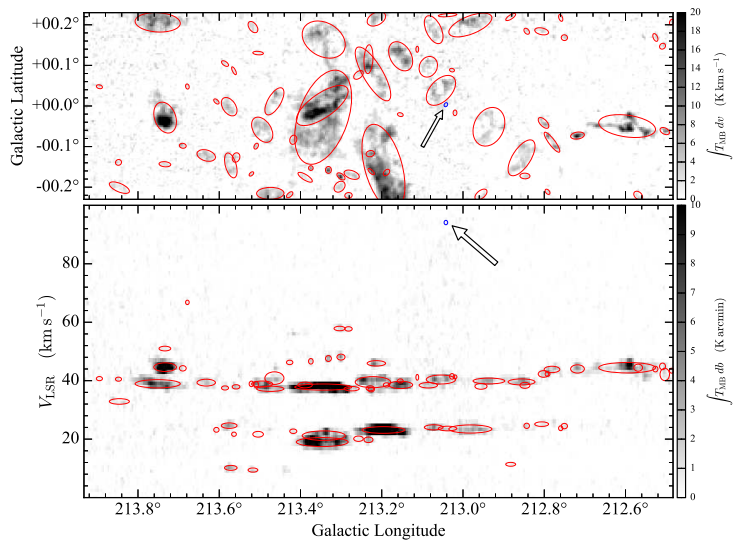}
 \end{center}
 \caption{
 Top:
 The Integrated intensity map obtained by integrating emission over the velocity range of 0 -- 100.1~km~s$^{-1}$.
 Bottom: 
 The $l$-$v$ diagram obtained by integrating the Galactic latitude range from $-$\timeform{0.25D} to $+$\timeform{0.25D}.
 The blue denotes G213.042$+$0.003 and the red ellipses denote the other cloud.
 The arrows point to the location of G213.042$+$0.003.
}
 \label{fig1}
\end{figure}

\section{Results}
\subsection{Integrated intensity and longitude-velocity map}
Figure \ref{fig1} shows the integrated intensity map and longitude-velocity ($l$-$v$) diagram.
The top panel shows the integrated intensity map obtained by integrating the emission over velocity range from 0 to 100.1~km~s$^{-1}$.
The bottom panel shows the $l$-$v$ diagram obtained by integrating the emission over the Galactic latitude range of $|b| < \timeform{0.25D}$.
Both figures were made by integrating only pixels exceeding 3$T_\mathrm{rms}$.
In the top panel, each ellipse indicates the size, ellipticity, and position angle of each identified cloud.
The position angle is determined using principle component analysis.
The rms spatial sizes along the major and minor axes are calculated by the intensity-weighted second moments.
The diameter sizes are estimated to multiply rms size by 2$\eta$, where $\eta =$ 1.91 \citep{2006PASP..118..590R}.
In the bottom panel, each ellipse shows the size and line width of each identified cloud.
The number of identified molecular clouds was 72 in this region.
The blue and red ellipses denote G213.042$+$0.003 and the others, respectively.

\begin{figure}[t] 
 \begin{center}
  \includegraphics[width=70mm]{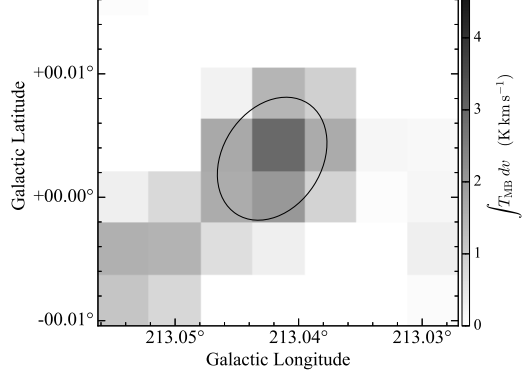}
 \end{center}
 \caption{
 The Integrated intensity map of G213.042$+$0.003 obtained by integrating emission over the velocity range of 93.6 -- 94.9~km~s$^{-1}$.
 Major and minor diameter of the ellipse were \timeform{38.4"} and \timeform{28.8"}, respectively, and the position angle was \timeform{57.2D}.
}
 \label{fig2}
\end{figure}

\begin{figure}
 \begin{center}
  \includegraphics[width=80mm]{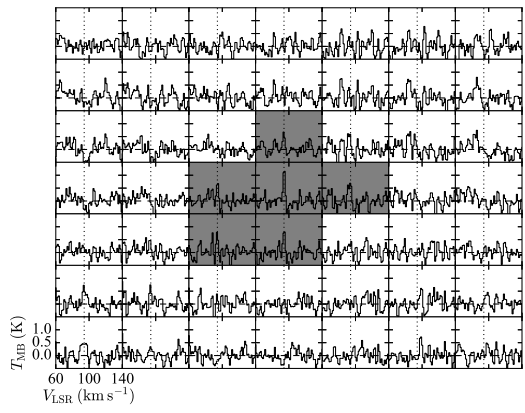}
 \end{center}
 \caption{
 The profile map of the same area as figure \ref{fig2}.
 The individual plots correspond to profiles of the individual pixels shown in figure \ref{fig2}.
 The vertical dotted line shown in each plot indicates the central velocity.
 The shaded spectra show the pixels of G213.042$+$0.003 identified by DENDROGRAM.
}
 \label{fig3}
\end{figure}

\subsection{G213.042$+$0.003}
In this section, we focus on G213.042$+$0.003 indicated with the blue ellipse in figure \ref{fig1}.

\subsubsection{Integrated intensity and profile map}
Figure \ref{fig2} shows the integrated intensity map of G213.042$+$0.003.
The solid line indicates an ellipse with major and minor diameters were \timeform{38.4"} and \timeform{28.8"}, respectively, and the position angle was \timeform{57.2D}.
Note that the size of molecular cloud was comparable to the beam size.
The center position of G213.042$+$0.003 was $(l, b) =$ (\timeform{213.042D}, \timeform{0.003D}) in the galactic coordinates and ($\alpha_\mathrm{J2000}$, $\delta_\mathrm{J2000}$) $=$ (\timeform{06h51m38.9s}, \timeform{-00d05m47.8s}) in the equatorial coordinates.
The error of the center position was \timeform{15.9"} which corresponds to half of the effective spatial resolution.
The center and FWHM line width of line profile at the central position were $V_\mathrm{LSR} =$ 94.1~km~s$^{-1}$ and $\Delta V =$ 1.5~km~s$^{-1}$, respectively.
The error of the center velocity was 0.8~km~s$^{-1}$ which corresponds to half of the velocity resolution.
We note that the FWHM line width was not resolved.
In table \ref{tb1}, the position and center velocity of G213.042$+$0.003 is summarized.

Figure \ref{fig3} shows the profile map, where position of each spectrum corresponds to each pixel shown in figure \ref{fig2}.
The spectra shaded in gray shows the pixels which were identified as G213.042$+$0.003.
The emission was detected at six pixels and the signal-to-noise ratio (S$/$N) at the peak was 4.9.

\begin{table}
%  \tbl{Heading of this tabular.}{%
% \begin{center}
  \caption{Position and center velocity of G213.042$+$0.003}
  \begin{tabular}{ll} \hline
    parameter & value \\ \hline
    Galactic longitude: $l$ & \timeform{213.042D} $\pm$ \timeform{0.004D} \\
    Galactic latitude: $b$ & \timeform{0.003D}  $\pm$ \timeform{0.004D} \\
    Right ascension: $\alpha_\mathrm{J2000}$ & \timeform{06h51m38.9s} $\pm$ \timeform{15.9s} \\
    Declination: $\delta_\mathrm{J2000}$ & \timeform{-00d05m47.8s}  $\pm$ \timeform{15.9s} \\
    LSR velocity: $V_\mathrm{LSR}$ & 94.1 $\pm$ 0.8 km s$^{-1}$ \\
    \hline
  \end{tabular}
  \label{tb1}
%  \begin{tabnote}
%    a brief note of table
%  \end{tabnote}
% \end{center}
\end{table}

\subsubsection{Kinematic distance}
\label{sec:distance}
The heliocentric distance of G213.042$+$0.003 was estimated to be $D =$ 21$_{-7}^{+12}$~kpc, which corresponded to the Galactocentric distance of $R =$ 29$_{-7}^{+12}$~kpc based on the kinematic distance adopting the solar Galactocentric distance of $R_0 =$ 8.34~kpc and the solar rotational velocity of  $V_0 =$ 240~km~s$^{-1}$, and assuming the derivative of the rotation curve beyond $R = R_0$ to be $dV/dR =$ $-$0.2~km~s$^{-1}$~kpc$^{-1}$ and the solar motion with respect to the LSR to be $(U_\odot, V_\odot, W_\odot) =$ (10.7, 15.6, 8.9)~km~s$^{-1}$ \citep{2014ApJ...783..130R}.

\citet{2014ApJ...783..130R} accurately obtained the Galactic constants and rotation curve to $R =$ 16~kpc based on the annual parallaxes measured by very long baseline interferometry.
Because the derivative of the rotation curve obtained to $R =$ 100~kpc by mass model was small \citep{1998MNRAS.294..429D}, it seems reasonable to adopt and extrapolate the result of \citet{2014ApJ...783..130R}.
The error of the kinematic distance is within 1~kpc due to the errors of the position and velocity.
In additon, the uncertainty of the kinematic distance is the range from $-$6~kpc to $+$11~kpc due to the errors of model by \citet{2014ApJ...783..130R}.
If IAU model ($R_0 =$ 8.5~kpc, $V_0 =$ 220~km~s$^{-1}$, and flat rotation curve)  was adopted, the kinematic distances were $D =$ 32~kpc and $R =$ 39~kpc.

\subsubsection{Size, luminosity, and mass}
The physical major and minor diameters were estimated to be 4.0$_{-1.3}^{+2.3}$~pc and 3.0$_{-1.0}^{+1.7}$~pc, respectively, based on the estimated heliocentric distance.
The \atom{H}{}{}$_2$ column density at the peak was estimated to be 1.2 $\times$ 10$^{21}$~\atom{H}{}{}$_2$~cm$^{-2}$ by integrating the brightness temperature $T_\mathrm{MB}$ over the velocity as expressed by $X_\mathrm{CO} \int T_\mathrm{MB} dv$ with the conversion factor $X_\mathrm{CO} = 5.6 \times 10^{20}~\mathrm{cm^{-2}~(K~km~s^{-1})^{-1}}$ \citep{2005PASJ...57..917N}.
Note that \atom{H}{}{}$_2$ column density may be underestimated because $X_\mathrm{CO}$ probably increases with the Galactocentric distance \citep{2010ApJ...710..133A}.
The \atom{C}{}{}\atom{O}{}{} luminosity mass was calculated as 2.5$_{-1.4}^{+3.7}$ $\times$ 10$^2$~$\Mo$ by multiplying $X_\mathrm{CO}$ and the proton mass, $m_\mathrm{p}$, with the \atom{C}{}{}\atom{O}{}{} luminosity, which was given by summing up the brightness temperature $T_\mathrm{MB}$ over all pixels and channels where \atom{C}{}{}\atom{O}{}{} emission was detected, as expressed with the equation $M_\mathrm{CO} = 2 \mu m_\mathrm{p} X_\mathrm{CO} \sum T_\mathrm{MB} dldbdv$, where $\mu =$ 1.36 considering the presence of helium.
The virial mass is estimated to be 2.2$_{-0.7}^{+1.3}$ $\times$ 10$^2$~$\Mo$ using the equation $M_\mathrm{VT} =$ 189$r_\mathrm{dc} \Delta V^2$, assuming the density profile of $\rho \propto r^{-1}$ \citep{2006PASP..118..590R}.
Table \ref{tb2} shows the physical parameters of G213.042$+$0.003.
Uncertainties are mainly come from the uncertainty of the distance to the object.

\begin{table}
%  \tbl{Heading of this tabular.}{%
% \begin{center}
 \caption{Physical parameters of G213.042$+$0.003}
  \begin{tabular}{ll} \hline
    parameter & value \\ \hline
    Galactocentric radius: $R$ & 29$_{-7}^{+12}$ kpc \\
    Distance from LSR: $D$ & 21$_{-7}^{+12}$ kpc \\
    Major diameter: $d_\mathrm{maj}$ & 4.0$_{-1.3}^{+2.3}$ pc \\
    Minor diameter: $d_\mathrm{min}$ & 3.0$_{-1.0}^{+1.7}$ pc \\
    Position angle & \timeform{57.2D} \\
    Spherical radius: $r$$^\ast$ & 1.7$_{-0.6}^{+1.0}$ pc \\
    Deconvolved radius: $r_\mathrm{dc}$$^\dagger$ & 0.52$_{-0.17}^{+0.30}$ pc \\
    FWHM line width: $\Delta V$ & 1.5 km s$^{-1}$ \\
    Peak temperature ($T_\mathrm{MB}$) & 1.15 K \\
    Integrated Intensity & 8.6 K km s$^{-1}$ \\
    \atom{C}{}{}\atom{O}{}{} luminosity: $L_\mathrm{CO}$ & 21$_{-12}^{+31}$ K km s$^{-1}$ pc$^2$ \\
    \atom{C}{}{}\atom{O}{}{} luminosity mass: $M_\mathrm{CO}$ & 2.5$_{-1.4}^{+3.7}$ $\times$ 10$^2$ $\Mo$ \\
    \atom{H}{}{}$_2$ column density: $N(\mathrm{H_2})$ & 1.2 $\times$ 10$^{21}$~\atom{H}{}{}$_2$~cm$^{-2}$ \\
    Virial mass: $M_\mathrm{VT}$ & 2.2$_{-0.7}^{+1.3}$ $\times$ 10$^2$ $\Mo$ \\
    \hline
  \end{tabular}
  \label{tb2}
  \begin{tabnote}
    $^\ast$ The spherical radius was calculated by $\frac{1}{2}\sqrt{d_\mathrm{maj}d_\mathrm{min}}$.  \\
    $^\dagger$ The deconvolved radius was calculated by $\sqrt{r^2 - r_\mathrm{beam}^2}$, where $r_\mathrm{beam}$ was the physical scale of beam radius.
  \end{tabnote}
% \end{center}
\end{table}

\begin{figure}
 \begin{center}
  \includegraphics[width=80mm]{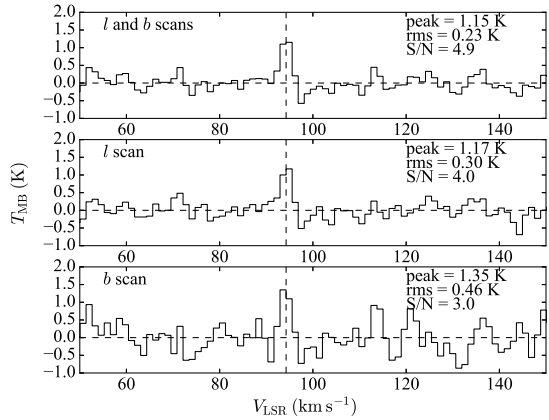}
 \end{center}
 \caption{
 The spectra of the peak position.
 The top panel shows the spectrum obtained by combining all scanned data, and the middle and bottom panels show spectra of scans in the $l$ and $b$ directions, respectively.
 The vertical dashed lines show $V_\mathrm{LSR} = $ 94.25~km~s$^{-1}$.
}
 \label{fig4}
\end{figure}

\section{Discussion}
\subsection{Possibility of spurious signal}
We divide a spectrum at the peak position of G213.042$+$0.003 into two spectra of individual scans in the Galactic longitude and latitude directions in order to check whether the detected emission is real and not made by a spurious signal.
The top panel of figure \ref{fig4} is the spectrum obtained by combining all the data, and the peak value is 1.15~K, the rms noise level is 0.23~K, and the resultant S$/$N is 4.9.
The middle and bottom panels show spectra of scans in the Galactic longitude and latitude directions, respectively.
Both spectra show that line emission is detected with the S$/$N of three or higher.
Therefore, we conclude that the emission is real and not spurious.

\subsection{Comparison with \atom{H}{}{}~{\sc i}}
\begin{figure}
 \begin{center}
  \includegraphics[width=100mm, bb=0 0 600 400]{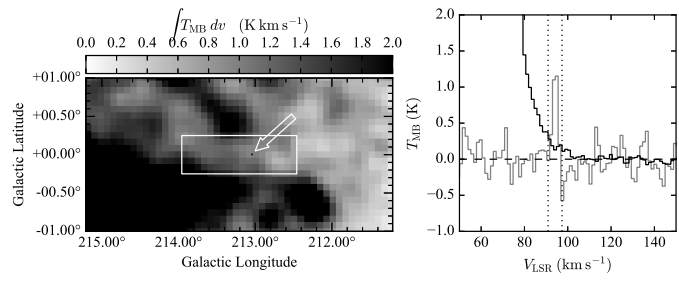}
 \end{center}
 \caption{
 Left:
 The Integrated intensity map of \atom{H}{}{}~{\sc i} (HI4PI) range of 91.6 -- 96.7~km~s$^{-1}$.
 The white rectangle shows the observed region in this study.
 The ellipse shows the position and size of G213.042$+$0.003.
 The arrows point to the location of G213.042$+$0.003.
 Right: 
 The black and gray lines are the \atom{H}{}{}~{\sc i} and \atom{C}{}{}\atom{O}{}{} spectra at the position of G213.042$+$0.003, respectively.
 $T_\mathrm{rms}$ of \atom{H}{}{}~{\sc i} spectrum was 0.034~K.
 The vertical dotted lines show the integrated velocity range of the left map.
}
 \label{fig5}
\end{figure}

We examine whether any \atom{H}{}{}~{\sc i} gas is associated with G213.042$+$0.003 using the \atom{H}{}{}~{\sc i} 4$\pi$ survey (HI4PI; \cite{2016A&A...594A.116H}) whose beam size is \timeform{972"}.
In figure \ref{fig5}, left panel shows the \atom{H}{}{}~{\sc i} map and that the G213.042$+$0.003,  which is indicated by a small ellipse, seems immersed in diffuse \atom{H}{}{}~{\sc i} envelope.
The right panel of figure \ref{fig5} shows \atom{H}{}{}~{\sc i} and \atom{C}{}{}\atom{O}{}{} spectra at the position of G213.042$+$0.003.
A weak \atom{H}{}{}~{\sc i} emission is seen at the same velocity range of the CO specctrum.
Further high angular resolution and sensitive \atom{H}{}{}~{\sc i} data are necessary to examine the structure in detail.

\subsection{Comparison with other molecular clouds}
We compare G213.042$+$0.003 with other molecular clouds located in the other outermost regions (\cite{1994ApJ...422...92D}, \cite{1994A&AS..103..503B}, \cite{2003ApJS..144...47B}, \cite{2001ApJ...551..852H}, and \cite{2015ApJ...798L..27S}).
We recalculate kinematic distances of these molecular clouds with the same parameters described in section \ref{sec:distance}.
The Galactocentric distance of Digel Cloud 2a \citep{1994ApJ...422...92D} is estimated as 23~kpc, and it is the most distant one among the clouds in previous studies.
Therefore, G213.042$+$0.003 ($R =$ 29~kpc) is the outermost cloud in the Galaxy at the present moment. 

\citet{2015ApJ...798L..27S} concluded that distant clouds identified by them were aligned with a new spiral arm, which was thought to be the extension of the Scutum-Crux arm.
If this spiral arm is extended to the third Galactic quadrant, its Galactocentric distance is 19~kpc in the direction of $l =$ \timeform{213D}, which is clearly different from the distance of G213.042$+$0.003.
Since there is no other molecular cloud around G213.042$+$0.003, it seems to be an isolated molecular cloud beyond the new arm.

G213.042$+$0.003 seems to be located in the region, where the surface densities of atomic and molecular gas are very low ($\Sigma_\mathrm{HI} <$ 1~$\Mo$~pc$^{-2}$ and $\Sigma_\mathrm{H_2} <$ 0.1~$\Mo$~pc$^{-2}$; \cite{2016PASJ...68....5N}).
Meanwhile, in external galaxies, no \atom{C}{}{}\atom{O}{}{} emission has been detected in the outermost regions with the sensitivity of 0.11~$\Mo$~pc$^{-2}$ with a 1~kpc beam \citep{2016MNRAS.455.1807W}.
Therefore, in order to detect \atom{C}{}{}\atom{O}{}{} emission and study molecular gas in outermost regions, higher resolution and sensitivity are necessary for external galaxies, and large-scale survey is necessary for the Galaxy.

\section{Summary}
We carried out mapping observations towards the outer Galactic region of $l =$ \timeform{212.5D} -- \timeform{214.0D} and $|b| <$ \timeform{0.25D} in the \atom{C}{}{12}\atom{O}{}{}($J =$ 1--0) line with the Nobeyama 45-m telescope and the multi-beam receiver system BEARS.
We detected a molecular cloud, G213.042$+$0.003, located at the Galactocentric distance of $R =$ 29$_{-7}^{+12}$~kpc, which is the molecular cloud whose kinematic distance is the largest  at the present moment.
The \atom{C}{}{}\atom{O}{}{} emission line of G213.042$+$0.003 is most likely a real but not spurious signal.

\begin{ack}
We thank anonymous referee and Dr. Masato Tsuboi for providing a thoughtful review.
We would like to thank all the staff of the Nobeyama Radio Observatory for supporting our observational project.
This research made use of Astropy, a community-developed core Python package for Astronomy (Astropy Collaboration, 2013, http://www.astropy.org).
This research made use of astrodendro, a Python package to compute dendrograms of Astronomical data (http://www.dendrograms.org/).
\end{ack}

\end{document}